\documentclass[nofootinbib,twocolumn,prd]{revtex4}
\usepackage{hyperref}
\usepackage{amsmath}
\usepackage{graphicx}
\usepackage{color}
\usepackage{ifpdf}

\newcommand\editremark[1]{ {\color{red} #1}}

\newcommand\hidetosubmit[1]{}
\newcommand\optional[1]{}
\newcommand\ForInternalReference[1]{}

\newcommand\Y[1]{Y^{(#1)}}

\newcommand\avL{\left< {\cal L}_{(a} {\cal L}_{b)} \right>}
\newcommand\avLh{\left\{ {\cal L}_{(a} {\cal L}_{b)} \right\}}
\newcommand\V{\hat{V}}
\newcommand\Vh{\hat{V}_h}
\newcommand\WeylScalar{{\psi_4}}

\newcommand\qmoperatorelement[3]{\left<#1\left|#2\right|#3\right>}

\def\bbh#1{binary black hole#1 (BBH#1)\gdef\bbh{BBH}}
\def\bh#1{black hole#1 (BH#1)\gdef\bh{BH}}

\begin{document}
\title{Asymptotic frame selection for binary black hole spacetimes II: Post-Newtonian limit} 
\author{E. Ochsner}
\email{evano@gravity.phys.uwm.edu}
\author{R.\ O'Shaughnessy}
\email{oshaughn@gravity.phys.uwm.edu}
\affiliation{Center for Gravitation and Cosmology, University of Wisconsin-Milwaukee,
Milwaukee, WI 53211, USA}

\begin{abstract}
One way to select a preferred frame from gravitational radiation is via the principal axes of  $\avL$, an average of the action of rotation group generators on the
Weyl tensor at asymptotic infinity.    In this paper we evaluate this time-domain average for a quasicircular binary using
approximate (post-Newtonian) waveforms.    For nonprecessing  unequal-mass binaries, we show the dominant eigenvector of
this tensor lies along the  orbital angular momentum.    For precessing binaries, this frame is not generally aligned
with either the orbital or total angular momentum, working to leading order in the spins.   
The difference between these two quantities grows with time, as the binary approaches the end of the inspiral and both
precession and higher harmonics become more significant.
\end{abstract}

\maketitle

\section{Introduction}
Ground-based gravitational wave detectors like LIGO and Virgo will soon detect the complicated gravitational-wave  signature of
merging binary compact objects.  
In general, all aspects of that signal will be significantly modulated by each compact binary's spin: 
misaligned spins break symmetry in the orbital plane, generally leading to strong modulations
in the inspiral \cite{ACST}, merger, and ringdown.
In the early stages, these  modulations are well-approximated 
 by quasistationary radiation from a circular orbit, slowly rotating and modulated 
as the orbital plane precesses
\cite{gwastro-mergers-nr-ComovingFrameExpansionSchmidt2010,ACST,gw-astro-mergers-approximations-SpinningPNHigherHarmonics,gw-astro-SpinAlignedLundgren-FragmentA-Theory}.
At late times,  numerical relativity simulations of mergers can efficiently predict gravitational radiation from a
variety of spins and orbits.   A well-chosen static or time-evolving frame makes it easier to compare their outputs to
one another and to these analytic models. 
Several methods to choose a frame have been proposed, using two broadly distinct algorithms.
One proposed method  chooses a frame such that the instantaneous $(l,m)=(2,2)$ emission is maximized
\cite{gwastro-mergers-nr-ComovingFrameExpansionSchmidt2010,gwastro-mergers-nr-Alignment-BoyleHarald-2011}.
Alternatively, a preferred frame follows by averaging a  tensor
constructed from the action of the rotation group on asymptotic emission  over
all orientations 
\cite{gwastro-mergers-nr-Alignment-ROS-Methods}.  
Though originally investigated numerically, both methods are analytically tractable and can be applied to existing
tables of  post-Newtonian waveforms' spin-weighted spherical harmonic modes \cite{gw-astro-mergers-approximations-SpinningPNHigherHarmonics}.
In this paper we evaluate $\avL$ in the post-Newtonian limit.

\section{Preferred orientations and quasicircular orbits}
\label{sec:Principles}

We explore gravitational wave emission from adiabatic quasicircular binary inspirals.   These orbits
can be parameterized by the two component masses $m_1,m_2$; component spins ${\bf S}_1,{\bf S}_2$; orbital separation
vector ${\bf r}$; and reduced orbital velocity ${\bf v} = \partial_t {\bf r}$.  
Following the notation of \cite{gw-astro-mergers-approximations-SpinningPNHigherHarmonics}, we perform our
post-Newtonian expansions using dimensionless mass and spin variables:
\begin{eqnarray}
\eta &\equiv& \frac{m_1 m_2}{(m_1+m_2)} \\ 
\delta &\equiv& \frac{m_1 - m_2}{m_1+m_2} \\
\mathbf{\chi}_\pm &=&\frac{1}{2}\left( {\bf S}_1/m_1^2 \pm {\bf S}_2/m_2^2\right) \\
\mathbf{\cal S} &=& \mathbf{\chi}_- + \delta \mathbf{\chi}_+ \\
 &=& \frac{1}{M} \left( \frac{{\bf S}_1}{m_1} - \frac{{\bf S}_2}{m_2} \right) \nonumber
\end{eqnarray}
We note that ${\cal S}$ is essentially a dimensionless version of the spin variable $\mathbf{\Delta}$ from~\cite{1995PhRvD..52..821K}, as the are related by $- M^2\,{\cal S} = \mathbf{\Delta}$. To describe the orbit and its orientation, we define the ``Newtonian'' angular momentum ${\bf L}_N = M\eta\,{\bf r} \times
\dot{{\bf r}}$; the unit separation vector $\hat{n} = \mathbf{r}/|\mathbf{r}|$; and the unit velocity vector 
$\hat{v} = \mathbf{v}/|\mathbf{v}|$.

The tensor $\avL$  is defined by averaging a certain  angular derivatives of the Weyl scalar $\psi_4$ over all
orientations, then normalizing the result  \cite{gwastro-mergers-nr-Alignment-ROS-Methods}.  Because this tensor is
constructed from rotation group generators ${\cal L}_a$, this normalized orientation average can be calculated efficiently  from
spin-weighted spherical harmonic basis coefficients $\WeylScalar_{lm}$:
\begin{eqnarray}
\label{eq:def:avL}
\avL &\equiv& 
 \frac{\int d\Omega \WeylScalar^*(t) {\cal L}_{(a}{\cal L}_{b)} \WeylScalar(t)
  }{
   \int d\Omega |\WeylScalar|^2
}
 \\
 &=& \frac{\sum_{lmm'} \WeylScalar_{lm'}^*  \WeylScalar_{lm}\qmoperatorelement{lm'}{{\cal L}_{(a}{\cal L}_{b)}}{lm} }{\int d\Omega |\WeylScalar|^2}  \nonumber
\end{eqnarray}
where in the second line we expand $\WeylScalar= \sum_{lm} \WeylScalar_{lm}(t)\Y{-2}_{lm}(\theta,\phi)$ and perform the angular integral.
In this way, the  average tensor $\avL$ at each time $t$ can be calculated algebraically; see the appendix for explicit formulae.
To better understand this average, we will contrast this expression with a similar average,  $\avLh$,  defined by replacing $\WeylScalar\rightarrow h$ in the above expression.

Several  general conclusions about the preferred orientations implied by $\avL$ follow from the definition.  
First and foremost, as gravitational radiation  is quadrupolar and predominantly driven by mass moments during inspiral, the tensor should
be well-approximated by the corresponding limit implied by instantaneous, purely quadrupolar emission  \cite{gwastro-mergers-nr-Alignment-ROS-Methods}:
\begin{eqnarray}
\avL &\simeq& \hat{L}_{Na} \hat{L}_{Nb} \left<m^2\right>  
\nonumber \\ &+&  \frac{\left<l(l+1)-m^2\right>}{2}  (\hat{n}_a \hat{n}_b + \hat{v}_a \hat{v}_b)
\end{eqnarray}
To the extent that the waveform is dominated by the leading-order $(l,|m|)=(2,2)$ modes, this tensor is diagonal, with
eigenvalues $(4,1,1)$.
The dominant eigendirection $\hat{V}$ of $\avL$ is therefore nearly the ``Newtonian'' orbital angular momentum.
Second, from perturbation theory, the dominant eigendirection $\hat{V}$ can be modified at leading order only through terms of the form $L_{a} X_b$ for some
vector $X$.
A post-Newtonian expansion of $\avL$ has additional convenient structure when $\avL$ is restricted to a constant-$l$
subspace.   
Specifically, in Eq. (\ref{eq:def:avL}), we employ terms in the numerator \emph{and denominator} from a single $l$.    
 Because only terms of constant $l$
enter, such an average has fixed trace:
\begin{eqnarray}
\text{Tr}\avL_l &=& l(l+1)
\end{eqnarray}
A series expansion of $\avL_l$ therefore consists of an expected zeroth-order term [${\cal T}^{(l)}$], followed by symmetric tracefree tensors [$A^{(lk)}$] of increasing
order and tightly constrained symmetry:
\begin{eqnarray}
\avL_l &\equiv & {\cal T}^{(l)}_{ab} + \sum_{k=0} A^{(lk)}_{ab} v^k \\
{\cal T}^{(l)}_{ab}&\equiv &[l^2 \hat{L}_{Na} \hat{L}_{Nb}  + l/2 (\hat{n}_a \hat{n}_b + \hat{v}_a \hat{v}_b)] \label{eq:tauL}
\end{eqnarray}
We seek to understand what tensors arise in a perturbation expansion and to determine how they impact the preferred
orientation $\hat{V}$.

\section{Rotation tensor in the post-Newtonian limit}
\label{sec:ApplyPN}

We employ a post-Newtonian model for the spin-weighted spherical harmonic modes of either the Weyl scalar, $\WeylScalar$, or the strain, $h$, for a generic precessing binary to construct the tensor $\avL$ or $\avLh$.
The modes $h_{lm}$ are provided in~\cite{gw-astro-mergers-approximations-SpinningPNHigherHarmonics} in a specific coordinate frame, where the $z$-axis is 
aligned with  $\bf{J}_0$, the total angular momentum at some initial time.  
In this frame, the instantaneous ``Newtonian'' orbital angular momentum $L_N=M\eta\,r\times \dot{r}$ has the form
\begin{eqnarray}
L_N &\equiv & \hat{z}\cos(\iota) + \sin\iota (\cos \alpha \hat{x} + \sin \alpha \hat{y})
\end{eqnarray}

We find that our expressions for the rotation tensor are simplest when given in a frame aligned with the Newtonian orbital angular momentum in which
\begin{eqnarray}
\hat{n} &=& \{ \cos \Phi\,,\,\sin \Phi\,,\, 0\} \\
\hat{v} &=& \{ - \sin \Phi \,,\, \cos \Phi\,,\, 0\} \\
\hat{L}_N &=& \{ 0\,,\, 0\,,\, 1\}\ .
\end{eqnarray}
The modes of~\cite{gw-astro-mergers-approximations-SpinningPNHigherHarmonics} will be given in this frame simply by evaluating the general expresions for $\iota=0$, $\alpha=\pi$. We derive our expressions for $\avL$ and $\avLh$ in this frame, but then express them in a frame-independent matter in terms of the vectors $\hat{n}$, $\hat{v}$ and $\hat{L}_N$. Thus, $\avL$ and $\avLh$ can be given in any frame simply by finding the components of $\hat{n}$, $\hat{v}$ and $\hat{L}_N$ in that frame.

In the non-precessing case, we find that the PN corrections to $\avL$ and $\avLh$ (and their restrictions to constant $l$ subspaces) can always be expressed as linear combinations of three symmetric, trace-free matrices. These matrices (and their components in the frame aligned with $\hat{L}_N$) are:
\begin{equation} \label{eq:STF1}
M^{(1)}_{ab} \equiv \hat{n}_a \hat{n}_b + \hat{v}_a \hat{v}_b = \left( \begin{matrix}
1 & 0 & 0\\
0 & 1 & 0\\
0 & 0 & 0
\end{matrix}\right)\ ,
\end{equation}
\begin{equation} \label{eq:STF2}
M^{(2)}_{ab} \equiv \hat{n}_a \hat{n}_b - \hat{v}_a \hat{v}_b = \left( \begin{matrix}
\cos 2 \Phi & \sin 2 \Phi & 0\\
\sin 2 \Phi & - \cos 2 \Phi & 0\\
0 & 0 & 0
\end{matrix}\right)\ ,
\end{equation}
\begin{equation} \label{eq:STF3}
M^{(3)}_{ab} \equiv \hat{L}_{Na}\,\hat{L}_{Nb} = \left( \begin{matrix}
0 & 0 & 0\\
0 & 0 & 0\\
0 & 0 & 1
\end{matrix}\right)\ .
\end{equation}

\subsection{Weyl scalar ($\WeylScalar$) in the Post-Newtonian quasistationary limit}
To evaluate $\avL$,
we convert from the
radiated strain moments $h_{lm}$ provided in \cite{gw-astro-mergers-approximations-SpinningPNHigherHarmonics} to $\psi_{lm}$ by taking  time derivatives:
\begin{eqnarray}
h_{lm} &=&h_{lm}(\Phi(t),v(t),\alpha(t),\iota(t),{\bf S}_1(t), {\bf S}_2(t)) \\
\partial_t h_{lm} &= & \frac{\partial h_{lm}}{\partial \Phi} \frac{\partial \Phi}{\partial t} +
 \frac{\partial h_{lm}}{\partial \alpha} \frac{\partial \alpha}{\partial t} +  
 \frac{\partial h_{lm}}{\partial \iota} \frac{\partial \iota}{\partial t} \\ 
\nonumber & { +}&  \frac{\partial h_{lm}}{\partial S_{1i}}\dot{\bf S}_{1,i}
 + \frac{\partial h_{lm}}{\partial S_{2,i}}\dot{\bf S}_{2,i} + \frac{\partial h_{lm}}{\partial v} \frac{\partial v}{\partial t}\\
\psi_{4,lm} &=& \partial_t^2 h_{lm} \hidetosubmit{\editremark{signs?}}
\end{eqnarray}
and organizing this expression self-consistently in powers of $v$.

For non-precessing binaries, all derivatives save $\dot{\Phi}$ and $\dot{v}$ are zero. At leading order, $\dot{\Phi} = \omega_{\rm orb} = v^3/M$ and $\dot{v} = (32\eta/5M)v^9$, with corrections to each of these known to 3.5PN relative order. Also note that the spin-independent terms in
$h_{lm}$ are known to $v^6$ (3PN)  \cite{2008CQGra..25p5003B,gwastro-pn-MultipoleMomentsNonspinning} beyond leading order and the spin-dependent terms to $v^4$ order (2 PN) \cite{gw-astro-mergers-approximations-SpinningPNHigherHarmonics}. Because $\dot{v}$ is order $v^6$ higher than $\dot{\Phi}$, $\dot{v}$ can be neglected so long as one is interested in expressions for $\psi_{lm}$ and $\avL$ accurate to less than 3PN order. In particular, we need to consider $\dot{v}$ only for 3PN-accurate $\psi_{lm}$ for non-spinning binaries.

For precessing binaries, the $h_{lm}$ are known only to 1.5PN order~\cite{gw-astro-mergers-approximations-SpinningPNHigherHarmonics}, so this is the highest order to which we can obtain $\psi_{lm}$ and $\avL$. 
Spin effects enter $h_{lm}$ at 1PN and 1.5PN order and $\dot{\Phi}$ at 1.5PN order. We must also account for the derivatives of $h_{lm}$ with respect to the spins, because $\dot{\bf S}_{i} \propto v^5$ which is 1PN order relative to $\dot{\Phi}$~\cite{WillWiseman:1996,gwastro-pn-MultipoleMomentsNonspinning}. Lastly, we must also keep track of $\dot{\iota}$ and $\dot{\alpha}$ (or equivalently $\dot{\hat{L}}_N$, as the angles $\iota$ and $\alpha$ describe the orientation of $\hat{L}_N$), since $\dot{\hat{L}}_N = \Omega_{L_N} \times \hat{L}_N \propto v^6$ (see Eq.~(\ref{eq:SpinsEvolve}) below) is 1.5PN order relative to $\dot{\Phi}$.

For geometrical reasons, the corrections from the derivatives $\dot{\alpha}$ and $\dot{\iota}$ depend only on $\Omega_{L_N}$, the vector around which $\hat{L}_N$ precesses, specifically as the combination
\begin{eqnarray}
\vec{\Upsilon} &=& \frac{M}{v^6}\,\hat{L}_N\times \partial_t \hat{L}_N \label{eq:Upsilon} \\
 &=&\frac{M}{v^6} \, \left(\Omega_{L_N} - (\Omega_{L_N}\cdot \hat{L}_N)  L_N\right)\ .\nonumber
\end{eqnarray}
Note that the motion of $\hat{L}_N$ creates a 1.5PN correction (relative to $\dot{\Phi}$) in $\psi_{lm}$, while we have defined $\Upsilon$ to be dimensionless by canceling out the $v^6/M$ dependence of $\Omega_{L_N}$.

\subsection{Rotation tensor for non-precessing binaries}
For a non-precessing binary, the tensors $\avL$ and $\avLh$ have the following frame-independent expressions:
\begin{widetext}
\begin{eqnarray}
\avL   &=& {\cal T}^{(2)} + \frac{\delta v^3}{16} ({\cal S}\cdot \hat{L}_N) \left[ M_{ab}^{(3)} - \frac{1}{2} M_{ab}^{(2)} - \frac{1}{2}  M_{ab}^{(1)} \right] 
  + \frac{v^2\delta^2}{2688} \left[ 40950 M_{ab}^{(3)} +129 \frac{1}{2} M_{ab}^{(2)} + 4129 \frac{1}{2}  M_{ab}^{(1)} \right] 
\label{eq:NPfirst}
   \\
\label{eq:avLh:Sum:Nonprecessing}
\avLh &=& {\cal T}^{(2)} + \delta v^3 ({\cal S}\cdot \hat{L}_N) \left[ M_{ab}^{(3)} - \frac{1}{2} M_{ab}^{(2)} - \frac{1}{2}  M_{ab}^{(1)} \right] 
  + \frac{v^2\delta^2}{168} \left[ 450 M_{ab}^{(3)} +39 \frac{1}{2} M_{ab}^{(2)} + 79 \frac{1}{2}  M_{ab}^{(1)} \right] 
\end{eqnarray}
\end{widetext}
Because gravitational wave emission is predominantly quadrupolar, these tensors are ${\cal T}^{(2)}$ at leading order, with PN corrections appearing at 1PN order.

We can also consider these tensors restricted to subsets of modes with constant $l$, consistently  in both the numerator
and denominator of the average.   To 1.5PN order, when we restrict these tensors to the $l=2,3,4$ subspaces we find
\begin{widetext}
\begin{align}
\avL_2 &= {\cal T}^{(2)} + \frac{v^2 \delta^2 -  3v^3 \delta ({\cal S} \cdot  \hat{L}_N) )}{96} \, \left( M_{ab}^{(1)} + M_{ab}^{(2)}  - 2 M_{ab}^{(3)} \right) \label{eq:NP2} \\
\avLh_2 &= {\cal T}^{(2)}  + \frac{v^2\delta^2  -3 v^3\delta ({\cal S} \cdot  \hat{L}_N) }{6}\, \left( M_{ab}^{(1)} + M_{ab}^{(2)} - 2 M_{ab}^{(3)}  \right) \label{eq:NPh2} \\
\avL_3 &= {\cal T}^{(3)}_{ab} + \left(\frac{303}{24\,604} + \frac{15\,(324\,043 - 1\,944\,258 \eta + 2\,585\,880 \eta^2)}{302\,678\,408(1-4\eta)}v^2 
\right)\,M^{(2)}_{ab}  
\nonumber \\
&+ \left( \frac{1}{24\,604} + \frac{15\,(1\,314\,379 - 7\,886\,274 \eta + 11\,827\,224
  \eta^2)}{302\,678\,408(1-4\eta)}v^2 
\right)\,(M^{(1)}_{ab} - 2 M_{ab}^{(3)}) \nonumber\\ 
&+ \frac{v^3 (\chi_{+} \cdot \hat{L}_{N})  \eta (1-3\eta)}{(1-4\eta)} \frac{19\,682\,880}{37\,834\,801} 
   (M^{(1)}_{ab} - 2 M_{ab}^{(3)} - \frac{1}{203} M^{(2)}_{ab})
\label{eq:NP3} \\
\avLh_3 &=  {\cal T}^{(3)}_{ab} + \frac{1}{304}[M^{(1)}_{ab} - 2 M_{ab}^{(3)} - 33 M^{(2)}_{ab}] + v^2 \frac{15}{46\, 208(1-4\eta)} 
\nonumber \\
 &\times \left[  \left(    (1039 - 6234 \eta + 9324 \eta^2)\right)\,(M^{(1)}_{ab} - 2 M^{(3)}_{ab})
 +     \left(391 - 2346 \eta + 3084 \eta^2)\,M^{(2)}_{ab}  \right) \right] \nonumber\\
 &+  \frac{v^3 (\chi_{+} \cdot \hat{L}_{N}) \eta(1-3\eta)}{(1-4\eta)}\frac{165}{1444}
   [23 (M^{(1)}_{ab} - 2 M_{ab}^{(3)}) -  M^{(2)}_{ab})] \label{eq:NPh3} 
\end{align}
\begin{align}
   \avL_4 &= {\cal T}^{(4)}_{ab} + \left( \frac{6}{7169} + \frac{23\,330\,181\,573 (1 - 2 \eta)^2
  \delta^2}{164\,462\,595\,200 (1 - 3 \eta)^2} v^2 \right)\, ( M^{(1)}_{ab}   -2 M_{ab}^{(3)} ) \nonumber\\
&+ \left( \frac{448}{7169} - \frac{66\,851\,487
  (1 - 2 \eta)^2 \delta^2}{32\,892\,519\,040 (1 - 3 \eta)^2} v^2 \right)\, M^{(2)}_{ab} \\
\avLh_4 &=  {\cal T}^{(4)}_{ab} + \left( \frac{6}{449} 
+ \frac{17\,982\,243 (1 - 2 \eta)^2 \delta^2}{40\,320\,200 (1 -  3 \eta)^2} v^2 \right)\, ( M^{(1)}_{ab}   -2 M_{ab}^{(3)} ) \nonumber\\
 &+ \left( \frac{112}{449} - \frac{141\,669 (1 - 2 \eta)^2 \delta^2}{8\,064\,040 (1 - 3 \eta)^2} v^2 \right)\, M^{(2)}_{ab}\ .  \label{eq:NPlast}
\end{align}
\end{widetext}

Note that in all cases these tensors are block diagonal, with a block corresponding to the orbital plane spanned by
$\hat{n}$ and $\hat{v}$ and a block for $\hat{L}_N$. In particular, there are no off-diagonal terms to couple the
contributions along $\hat{L}_N$ to those in the orbital plane. This means that the dominant eigenvector will remain along $\hat{L}_N$ at all times with an eigenvalue $\sim l^2$ and there will be two
eigenvectors in the orbital plane with eigenvalues $l/2$. While the \emph{eigenvectors} will be constant, the PN
corrections will introduce small, time-dependent variations in the \emph{eigenvalues}. 
In fact, for nonprecessing binaries this direction is tightly enforced by symmetry and cannot be
modified, for example, by omitted nonlinear memory terms; see Appendix~\ref{appA} for details.

The tensors extracted from constant-$l$ subspaces can behave pathologically when multipoles of that order are strongly
suppressed by symmetry.  
In particular, the $(3,3)$ and $(3,1)$ modes vanish in the equal mass case where $\delta = 0$. As a result, the
expressions for $\avL_3$ and $\avLh_3$ will diverge in the limit $\delta \rightarrow 0$ because they contain a factor
$(1 - 4\eta) = \delta^2$ in the denominator.  

\subsection{Rotation tensor for precessing binaries}

In the precessing case, when the spins are not aligned with the orbital angular momentum, the rotation tensors are no
longer block-diagonal. These off-diagonal terms depend on the transverse components of spin and mix contributions in the
orbital plane with those along $\hat{L}_N$. As a result, the eigendirections will vary in time as the binary precesses
and they will not lie along $\hat{L}_N$ and in orbital plane. 

As in the non-precessing case, we derive the expressions for $\avL$ and $\avLh$ in a frame aligned with the
instantaneous $\hat{L}_N$ but express them in a frame-independent way so that they can be used in any frame.%
 As in the
non-precessing case, all of the tensors can be expressed as linear combinations of a small collection of matrices. The
new matrices needed for the precessing case (and their components in the frame aligned with the instantaneous
$\hat{L}_N$) are: 
\begin{equation} \label{eq:STF4}
M^{(1)}_{ac} {\cal S}_c \hat{L}_{Nb} = \left( \begin{matrix}
0 & 0 & \frac{{\cal S}_x}{2}\\
0 & 0 & \frac{{\cal S}_y}{2}\\
\frac{{\cal S}_x}{2} & \frac{{\cal S}_y}{2} & 0
\end{matrix}\right)\ ,
\end{equation}
\begin{widetext}
\begin{equation} \label{eq:STF5}
M^{(2)}_{ac} {\cal S}_c \hat{L}_{Nb} = \left( \begin{matrix}
0 & 0 & \frac{{\cal S}_x}{2} \cos 2\Phi + \frac{{\cal S}_y}{2} \sin 2 \Phi\\
0 & 0 & \frac{{\cal S}_x}{2} \sin 2\Phi - \frac{{\cal S}_y}{2} \cos 2 \Phi\\
\frac{{\cal S}_x}{2} \cos 2\Phi + \frac{{\cal S}_y}{2} \sin 2 \Phi & \frac{{\cal S}_x}{2} \sin 2\Phi - \frac{{\cal S}_y}{2} \cos 2 \Phi & 0
\end{matrix}\right)
\end{equation}
\end{widetext}
and also $M^{(1)}_{ac} \chi_{+c} \hat{L}_{Nb}$ and $M^{(2)}_{ac} \chi_{+c} \hat{L}_{Nb}$, which are the equivalent to the expressions above, but with the components of $\chi_+$ replacing those of ${\cal S}$.

The $\avL$ and $\avLh$ tensors and their constant $l$ restrictions for precessing binaries are given by 
\begin{widetext}
\begin{eqnarray}
\avL^{\rm prec} &=& \avL^{\rm NP} + v^3 \left[  M_{ac}^{(1)}
 \left( \frac{581}{96}\delta\, {\cal S} + 29 \eta\, \chi_+ \right)_c \hat{L}_{Nb)}  + \frac{\delta}{96} M^{(2)}_{(ac} {\cal S}_c \hat{L}_{Nb)}\right] \label{eq:Pfirst} \nonumber\\
 &+& 6\,v^3\, \hat{L}_{N(a} \Upsilon_{b)} \\
\label{eq:avLh:Sum:Delta:Precessing}
\avLh^{\rm prec} &=& \avLh^{\rm NP} + v^3 \left[  M_{ac}^{(1)}
\left( \frac{41}{6}\delta\, {\cal S} + 29\eta\, \chi_+ \right)_c \hat{L}_{Nb)} 
-  M^{(2)}_{(ac}  \left(   \frac{35}{6} \delta\, {\cal S}_c  + 21 \eta\, \chi_{+c} \right)  \hat{L}_{Nb)} \right] \label{eq:Ph}\\
\avL_2^{\rm prec} &=& \avL_2^{\rm NP} + v^3 \left[  M_{ac}^{(1)}
 \left( \frac{581}{96}\delta\, {\cal S} + 29 \eta\, \chi_+ \right)_c \hat{L}_{Nb)}  + \frac{\delta}{96} M^{(2)}_{(ac} {\cal S}_c \hat{L}_{Nb)}\right] \nonumber\\
 &+& 6\,v^3\, \hat{L}_{N(a} \Upsilon_{b)} \\
\avLh_2^{\rm prec} &=& \avLh_2^{\rm NP} + v^3 \left[  M_{ac}^{(1)}
\left( \frac{41}{6}\delta\, {\cal S} + 29\eta\, \chi_+ \right)_c \hat{L}_{Nb)} 
-  M^{(2)}_{(ac}  \left(   \frac{35}{6} \delta\, {\cal S}_c  + 21 \eta\, \chi_{+c} \right)  \hat{L}_{Nb)} \right] \\
\avL_3^{\rm prec} &=& \avL_3^{\rm NP} -   v^3 \frac{7680 (1-3\eta)}{6151(1-4\eta)} M^{(1)}_{(ac} \chi_{+c} \hat{L}_{Nb)} \\
\avLh_3^{\rm prec} &=& \avLh_3^{\rm NP}   - v^3 \frac{120 (1-3\eta)}{19(1-4\eta)} M^{(1)}_{(ac} \chi_{+c} \hat{L}_{Nb)} \label{eq:Ph3} \\
\avL_4^{\rm prec} &=& \avL_4^{\rm NP} \\
\avL_4^{\rm prec} &=& \avL_4^{\rm NP} \label{eq:Plast}
\end{eqnarray}
\end{widetext}

At high multipole order, the accuracy of these expressions is inevitably limited by the accuracy of post-Newtonian
expansions, which contain terms explicitly coupling to spins only to modes $l\le 3$
\cite{gw-astro-mergers-approximations-SpinningPNHigherHarmonics,WillWiseman:1996}.
Hence, the $l=4$ subspace does not have any off-diagonal terms through 1.5PN order, though it will presumably acquire
such terms at a higher PN order.

By examining Eq. (4.17) of~\cite{gw-astro-mergers-approximations-SpinningPNHigherHarmonics}, we can understand the origin of these off-diagonal terms with transverse spins. It may be surprising that they do not appear until 1.5PN order, since $h_{22}$ has such terms already at 1PN order. However, as explained in Appendix~\ref{appA}, these transverse terms are the result of coupling modes with $m$ values that differ by $1$. As a result, the 1PN transverse spin terms in $h_{22}$ must couple with the leading 0.5PN order term in $h_{21}$, so that the off block-diagonal contribution to $\avLh$ is 1.5PN order. These terms in $h_{22}$ and $h_{21}$ are both proportional to the first harmonic of the orbital phase, so this orbital scale dependence cancels out and creates a contribution to $\avLh$ of the form of Eq.~(\ref{eq:STF4}) (i.e. $M^{(1)}_{ac} S_c \hat{L}_{Nb}$ for some combination of spins $S$). There are also 1.5PN order transverse spin terms in $h_{21}$ proportional to the second and zeroth harmonic that couple to the leading order term in $h_{22}$. These create contributions to $\avLh$ of the form of Eqs.~(\ref{eq:STF4}) ($M^{(1)}_{ac} S_c \hat{L}_{Nb}$ ) and~(\ref{eq:STF5}) ($M^{(2)}_{ac} S_c \hat{L}_{Nb}$), respectively.

In a similar manner, $h_{33}$ and $h_{31}$ have transverse spin terms proportional to the second harmonic which couple to the leading term in $h_{32}$ and create off-diagonal terms in $\avLh$ of the form of Eq.~(\ref{eq:STF4}). These terms are 2.5PN order, and so they do not appear in Eq.~(\ref{eq:Ph}). However, because the leading $l=3$ terms are 0.5PN order, and in computing $\avLh_3$ we restrict both the numerator \emph{and} the denominator to $l=3$, these terms are 1.5PN relative to the leading order of $\avLh_3$ and therefore appear at that order in Eq.~(\ref{eq:Ph3}). The $l=4$ modes do not have transverse spin terms through 1.5PN order, and so we do not see any off-diagonal terms in $\avLh_4$.

\subsection{Broken symmetry from precession}

For a precessing binary at 1.5PN order, $\avL$ no longer has the Newtonian, orbital, 
or even total angular momentum as an eigenvector.   
The orbital and  total angular momenta are given by the spin-dependent expressions
\cite{1995PhRvD..52..821K,2006PhRvD..74j4033F,2005PhRvD..71h4027W}: %
\begin{subequations}
\label{eq:def:LJ}
\begin{eqnarray}
{\bf J}&=& {\bf L}+{\bf S}_1 + {\bf S}_2 \\
{\bf L}_N &=& M \eta {\bf r} \times {\bf v} \\
{\bf L}_{PN}&\simeq&  {\bf L}_N \frac{7-\eta}{2} v^2 \\
({\bf L}_{SO})_a &= &M^2 \eta v^2 \epsilon_{abc}\epsilon_{bpq} 
 \\ &\times &
\ [ n^b n_p (\delta \chi_{-q}+ (1-\eta)\chi_{+q}) - v_b v_p \eta \chi_{+q}] \nonumber \\
{\bf L}&=&  {\bf L}_N +{\bf L}_{PN} + {\bf L}_{SO} \\
 &=& M^2 \eta  v^{-1} \hat{L}_N \left( 1 + \left( \frac{3}{2} + \frac{\eta}{6} \right)v^2 \right. \nonumber\\ 
 &+& \left. \left( 2 \delta {\cal  S}  -\frac{10}{3} \left( (1 - 2 \eta) \chi_s + \delta \chi_a \right) \right) \cdot \hat{L}_N  v^3 \right) \nonumber\\
 &+& M^2 \eta  v^2 \left[ \hat{n} \times \left( \hat{n} \times \left( 3\left( (1 - 2 \eta) \chi_s + \delta \chi_a \right) - \delta  {\cal S} \right)  \right) \right.\nonumber\\
 &+& \left. \hat{v} \times \left( \hat{v} \times \left( -\frac{1}{2}\left( (1 - 2 \eta) \chi_s + \delta \chi_a \right) + \frac{\delta}{2} {\cal S} \right) \right) \right] \nonumber
\end{eqnarray}
\end{subequations}
where we have explicitly expanded ${\bf L}$ to 1.5PN order.
Because the Newtonian angular momentum at leading order is $  \propto v^{-1}$, the term  $L_{SO}$ in this expansion is $v^3$ beyond leading
order.  %

At and below 1PN order $\hat{L} = \hat{L}_N$. The 1.5PN order spin-orbit contribution to the orbital angular momentum,
${\bf L}_{SO}$ breaks this alignment and introduces a component in the instantaneous orbital plane that oscillates at twice the orbital frequency.
Likewise, the spin-dependent terms at order $v^3$ in $\avL$ and $\avLh$ are not block diagonal and  time
dependent.  
As we will show by concrete example in Section \ref{sec:NumericalExamples}, these two time-dependent expressions do not
  conspire to evolve together: ${\bf L}$ is not an eigenvector of either $\avL$, $\avLh$, or most (but not all) of the
  tensors produced from constant-$l$ subspaces.
To demonstrate this explicitly, one can laboriously evaluate the expressions
\begin{eqnarray}
{\bf X}_p^l \equiv \epsilon_{paq} \avL_l L^b L^q \\
{\bf Y}_p^l \equiv \epsilon_{paq}\avL_l J^b J^q \\
{\bf Z}_p^l \equiv \epsilon_{paq} \avL_l L_N^b L_N^q 
\end{eqnarray}
An eigenvector $X^a$ of the three-dimensional matrix $A_{ab}$ necessarily has  $\epsilon_{abc}X^bA_{cd}X^d=\lambda
\epsilon_{abc}X^bX^c=0$.  
For $\avL$ and $\avLh$, we find each of these vectors are nonzero at1.5PN order, once the off-diagonal precession terms enter.

\subsection{Memory terms}
In the calculations above, we use the ``ready to use'' waveforms provided by
\citet{gw-astro-mergers-approximations-SpinningPNHigherHarmonics}.  These expressions do not  include nonlinear memory
terms, which enter at leading PN order and beyond.   In terms of the modes used in this paper, these terms
would produce a strain multipole $h_{20}$ comparable in magnitude to $h_{22}$ for nonprecessing binaries  \cite{1991PhRvD..44.2945W,2010CQGra..27h4036F}.

No complete expression exists in the literature for nonlinear memory for generic spinning systems.   Memory terms
depend on the integrated past history of the binary; they therefore can differ in magnitude, angular dependence, and functional form
versus time depending on how the spins have evolved in the past.
That said, we anticipate our estimate of this particular preferred orientation (i.e., the dominant eigenvector of
$\avL$) will not change significantly when memory is included.
For example,  because memory  terms cannot modify the reflection symmetry  for nonprecessing binaries, they likewise
cannot change the preferred orientation away from the normal to the orbital plane.   

\section{Numerical evaluation and discussion of results }
\label{sec:NumericalExamples}
\begin{figure}
\includegraphics[width=\columnwidth]{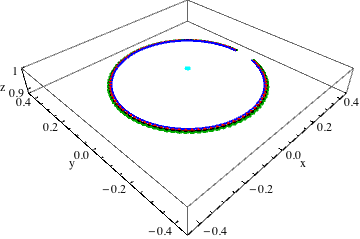}
\includegraphics[width=\columnwidth]{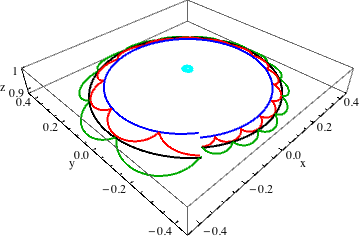}
\caption{\label{fig:Evan:Trajectories:BHNS}\textbf{Comparing trajectories 1: BH-NS binary}:  These panels show the trajectories of $\hat{L}_N$ (black), $\hat{L}$ (green), $\hat{J}$ (cyan), $\hat{V}$ the principal eigendirection of $\avL$ (blue) and $\hat{V}_h$ the principal eigendirection of $\avLh$ (red) for a  $10+1.4M_\odot$ BH-NS binary where the black hole has maximal spin tilted $30^\circ$ relative to the Newtonian orbital angular momentum at an initial frequency of 40 Hz. The frame is chosen so that the $z$-axis coincides with the initial value of $\hat{J}$. The top panel shows the first precessional cycle, which spans a frequency range of roughly $40 - 47$ Hz. The bottom panel shows a very late precessional cycle shortly before merger would occur, covering a frequency band of $200 - 600$ Hz.
}
\end{figure}

\begin{figure}
\includegraphics[width=\columnwidth]{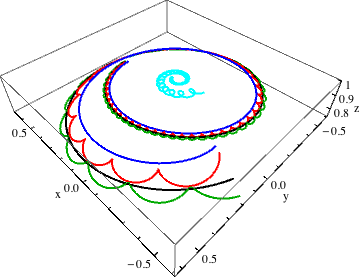}
\caption{\label{fig:Evan:Trajectories:BBH}\textbf{Comparing trajectories 2: BH-BH binary}:  This plot is the same as Fig.~\ref{fig:Evan:Trajectories:BHNS}, but for a $60+30M_\odot$ BH-BH binary with both spins maximal, perpendicular to the Newtonian orbital angular momentum and perpendicular to each other at an initial frequency of 10 Hz. This plot shows the entire inspiral of the binary, reaching a GW frequency of about 85 Hz at the end.}
\end{figure}

Our calculations above have demonstrated that $\hat{V}$ and $\hat{V}_h$, the principal eigendirections of $\avL$ and $\avLh$,  differ from
$\hat{L}$, $\hat{J}$, and
and $\hat{L}_N$, with a difference that grows with time. To illustrate this difference, we have simulated the
evolution of two  precessing, inspiralling binaries: a $10+1.4 M_\odot$ BH-NS binary where the black hole has a maximal spin tilted $30^\circ$ relative to the Newtonian orbital angular momentum at its initial frequency of 40 Hz and a $60+30 M_\odot$ BH-BH binary where both black hole spins are maximal and in the orbital plane with a $90^\circ$ angle between the two spins at an initial frequency of 10 Hz. Figs.~\ref{fig:Evan:Trajectories:BHNS} and~\ref{fig:Evan:Trajectories:BBH} plot the trajectories of $\hat{L}_N$, $\hat{L}$, $\hat{J}$, $\hat{V}$ and $\hat{V}_h$ for these two binaries.

The equations of motion were numerically integrated using the  \texttt{lalsimulation}\footnote{The \texttt{lalsimulation} package encapsulates and standardizes several waveform generation codes; see
  the LIGO Algorithms Library \cite{LAL} for more details.} 
implementation of the adiabatic, spinning Taylor T4 approximation described in
\cite{BCV:PTF}.   
Specifically, this code evolves the binary phase and frequency, spin and Newtonian orbital angular momentum vectors equations according to
[see, e.g., \cite{ACST,BCV:PTF,2004PhRvD..70l4020S}]: %
\begin{subequations}
\label{eq:SpinsEvolve}
\begin{align}
\partial_t \Phi &= \frac{v^3}{M} \\
\partial_t v &= - \frac{\cal F}{M dE/dv} \\
\partial_t \hat{L}_N &= \vec{\Omega}_{L_N} \times \hat{L}_N \label{eq:prec_eq} \\
\partial_t \vec{S}_1 &= \vec{\Omega}_{S_1} \times \vec{S}_1 \\
\partial_t \vec{S}_2 &= \vec{\Omega}_{S_2} \times \vec{S}_2 \\
\vec{\Omega}_{L_N} &= \frac{v^6}{M} \Bigg [  
  (2+ \frac{3 m_2}{2 m_1}) \frac{\vec{S}_1}{M^2}  + (2+ \frac{3 m_1}{2 m_2}) \frac{\vec{S}_2}{M^2} \nonumber \\  
  & ~ - \frac{3 v}{2 M^2 \eta} [ (\vec{S}_2 \cdot \hat{L}_N)\vec{S}_1 +  (\vec{S}_1 \cdot \hat{L}_N)\vec{S}_2] \Bigg] \\
\vec{\Omega}_{S_1} &= \frac{v^5}{M}\left[  \eta
  (2+ \frac{3 m_2}{2 m_1}) \hat{L}_N + v\,\frac{\vec{S}_{2}-3\hat{L}_N (\vec{S}_2\cdot \hat{L}_N)}{2 M^2}
  \right] \\
  \vec{\Omega}_{S_2} &= \frac{v^5}{M}\left[  \eta
  (2+ \frac{3 m_1}{2 m_2}) \hat{L}_N + v\,\frac{\vec{S}_{1}-3\hat{L}_N (\vec{S}_1\cdot \hat{L}_N)}{2 M^2}
  \right]\label{eq:SpinsEvolveLast}
\end{align}
\end{subequations}
where $E$ and ${\cal F}$ are the PN-expanded binary energy and gravitational-wave flux (with spin contributions included), respectively. ${\cal F}/(M dE/dv)$ is then re-expanded as a polynomial in $v$, as described in~\cite{2009PhRvD..80h4043B}.
At each time, we obtain values for $\avL$ and $\avLh$ by taking the dynamical variables computed via Eqs.~(\ref{eq:SpinsEvolve}) and plugging these into Eqs.~(\ref{eq:Pfirst})-(\ref{eq:Plast}). We then compute the principal eigenvectors and eigenvalues of $\avL$ and $\avLh$ (and their constant-$l$ subspaces), which are used to create the plots in this section.

\begin{figure*}
\includegraphics[width=\columnwidth]{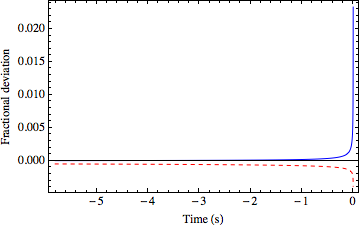}
\includegraphics[width=\columnwidth]{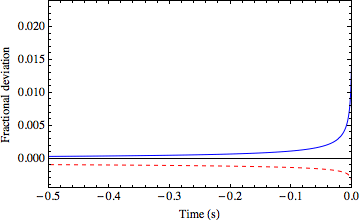}
\includegraphics[width=\columnwidth]{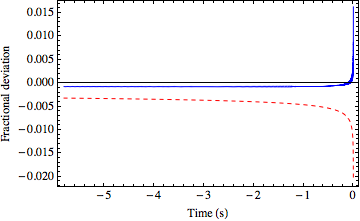}
\includegraphics[width=\columnwidth]{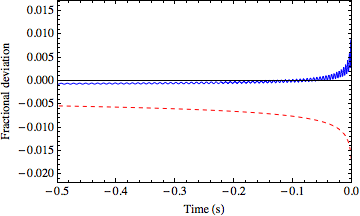}
\caption{\label{fig:Evan:Eigenvalues}\textbf{Comparing eigenvalues}:  The fractional deviation $(\lambda/\bar{\lambda}-1)$ in the
  dominant eigenvalues (where $\bar{\lambda}$ is the unperturbed eigenvalue and $\lambda$ is the actual eigenvalue) of $\avL_l$ (top panels) and $\avLh_l$ (bottom panels), for the BH-NS binary orbit described in Figure
  \ref{fig:Evan:Trajectories:BHNS}.  The blue curve indicates $l=2$; the red dashed curve shows $l=3$.
  For reference, a black solid line is shown at $0$.  
\emph{Left panels}:  Eigenvalues of $\avL$ (top) and $\avLh$ (bottom) for the full evolution shown.
\emph{Right panels}: As above, but limited to a short interval immediately preceding the last simulated post-Newtonian
orbit, where the minimum-energy condition breaks down.
}
\end{figure*}

\subsection{Orientations}
The trajectories shown in Figures \ref{fig:Evan:Trajectories:BHNS} and \ref{fig:Evan:Trajectories:BBH} demonstrate that,
for these two prototype binaries, the four directions $\hat{L},\hat{L}_N,\V$ and $\Vh$ generally do not coincide.  
Though these four orientations have very roughly the same features -- all precess  -- all four exhibit distinctly
different trends.
For example, the Newtonian angular momentum $\hat{L}_N$ (black) precesses smoothly for our analytic model.   The orbital angular
momentum $\hat{L}$ oscillates (nutates) around $\hat{L}_N$.   
The preferred orientation $\V$ extracted from the Weyl scalar (blue) also precesses smoothly in our coordinates,
tracking with but offset somewhat from $\hat{L}_N$.    Finally, the preferred orientation $\Vh$ extracted from the
strain (red) oscillates around $\V$.

As expected from their functional form -- a power series in $v$ -- the differences between these four quantities grow
more pronounced as $v$ grows, closer to the merger event.   
As a consequence, in the fixed sensitive band of ground-based detectors, higher-mass sources exhibit more pronounced
differences between these four orientations.   
We cannot reliably extrapolate to the late stages of merger.  However, as suggested by previous studies
\cite{gwastro-mergers-nr-Alignment-ROS-IsJEnough}, we expect substantial differences can accumulate between these orientations.

Among these four orientations, only $\vec{L}_N$ and $\V$ evolve smoothly, without exhibiting strong oscillations on the
orbital period.   By contrast, the direction $\Vh$ extracted from the strain oscillates significantly on the orbital period.
The differing behavior of  $\Vh$ and $\V$ can be understood entirely from the magnitudes of the terms of the form of Eq.~(\ref{eq:STF4}) and Eq.~(\ref{eq:STF5}) which appear in Eqs.~(\ref{eq:Pfirst}) and (\ref{eq:Ph}). Note that the former depend only on the spin components, and so create smooth perturbations to the eigenvectors which vary on the precessional time scale. The latter vary as twice the orbital phase, and so create perturbations that oscillate much more rapidly.

So, by comparing Eqs.~(\ref{eq:Pfirst}) and (\ref{eq:Ph}) we see that the coefficients of 
$M^{(1)}_{ab}$ in $\avL$ and $\avLh$ have very similar magnitudes, and so offset $\V$ and $\Vh$ from $\hat{L}_N$ by similar amounts. However, the coefficient multiplying $M^{(2)}_{ab}$ in $\avL$ is always much smaller than the coefficient multiplying $M^{(1)}_{ab}$. Therefore, these oscillations are strongly suppressed and are not apparent in Figures \ref{fig:Evan:Trajectories:BHNS} and \ref{fig:Evan:Trajectories:BBH}. On the other hand,  the coefficient multiplying $M^{(2)}_{ab}$ in $\avLh$ is nearly as large as the coefficient multiplying $M^{(1)}_{ab}$. Therefore, the orbital-scale oscillations are much more significant and creates the distinct ``crown'' shape in the trajectory of $\Vh$ in Figures \ref{fig:Evan:Trajectories:BHNS} and \ref{fig:Evan:Trajectories:BBH}.

Another difference between $\Vh$ and $\V$ is the presence of the $\hat{L}_{N(a}\Upsilon_{b)}$ term in Eq.~(\ref{eq:Pfirst}) that arises from the time derivatives of $\hat{L}_N$, where $\Upsilon$ is given by Eq.~(\ref{eq:Upsilon}). 
One can see that $\Upsilon$ will always point back towards the center of the precessional cone ($\hat{J}$), so that this term always tends to shrink the precessional cone of $\V$ relative to $\Vh$. 
In particular, in Figs.~\ref{fig:Evan:Trajectories:BHNS} and~\ref{fig:Evan:Trajectories:BBH} we see that the trajectory of $\V$ lies just above the oscillations of the trajectory of $\Vh$ and we have checked that if the  $\hat{L}_{N(a}\Upsilon_{b)}$ term were removed $\V$ would pass through the center of the oscillations of $\Vh$.

Let us consider $\V$ and $\Vh$ in the equal mass limit ($\delta \rightarrow 0$, $\eta \rightarrow 1/4$). From Eqs.~(\ref{eq:Pfirst}) and (\ref{eq:Ph}) it is clear the off-diagonal terms will vanish if and only if $\chi_+ = 0$, which means the spins must be equal and opposite. But in this case the total spin vanishes and the binary does not precess. On the other hand, in the extreme mass ratio limit ($\delta \rightarrow 1$, $\eta \rightarrow 0$) the off-diagonal terms will vanish if and only if ${\cal S} = 0$. But this can only happen if $\mathbf{S}_1 / m_1 = \mathbf{S}_2 / m_2$ which would mean that the dimensionless spin of the larger body must be extremely tiny and the system will again not precess significantly. In intermediate cases, the off-diagonal terms would vanish only if the contributions from the $\chi_+$ symmetric spin terms and the ${\cal S}$ antisymmetric spin terms were carefully tuned to cancel out. However, one cannot simultaneously cancel the terms multiplying both $M^{(1)}_{ab}$ and $M^{(2)}_{ab}$ in  Eqs.~(\ref{eq:Pfirst}) and (\ref{eq:Ph}). Therefore, we can conclude that for \emph{any} precessing binary the dominant eigendirections $\V$ and $\Vh$ cannot coincide with $\hat{L}_N$.

\begin{figure}
\includegraphics[width=\columnwidth]{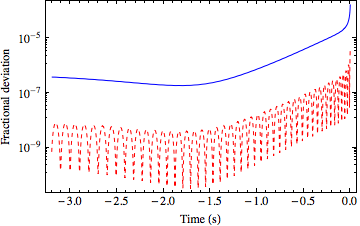}
\includegraphics[width=\columnwidth]{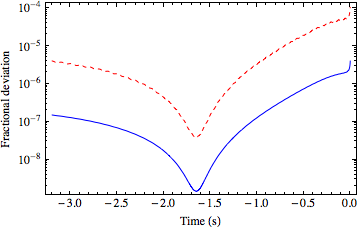}
\caption{\label{fig:Evan:SubspaceDirectionsCorrespond}\textbf{Orientations of subspace eigenvectors}: 
\emph{Top panel}: A plot of $1-\V_2\cdot \V$ (solid blue) and $1-\hat{V}_{h2} \cdot \Vh$ (dashed red) 
versus time for the BH-BH binary orbit described in Figure~\ref{fig:Evan:Trajectories:BBH}.
\emph{Bottom panel}: A plot of $1-\V_3 \cdot \hat{L}_N$ (solid blue) and 
$1- \hat{V}_{h3} \cdot \hat{L}_N$ (dashed red) versus time for the same BH-BH binary. 
The eigenvectors from the $l=2$ ($l=3$)
subspaces are rather close to the eigenvector of the full rotation tensor ($\hat{L}_N$).}
\end{figure}

To this point we have described the preferred orientations $\V$ and $\Vh$ associated with $\avL$ and $\avLh$, summing over the
contribution from all subspaces.  We can also define the two preferred orientations associated with the constant-$l$ subspaces of $\avL$ and $\avLh$.   One would expect (and Figure
\ref{fig:Evan:SubspaceDirectionsCorrespond} confirms) that, because leading-order quadrupole emission dominates the
gravitational wave signal, $\V_2$ ($\hat{V}_{h2}$) would be extremely close to $\V$ ($\Vh$).  
By contrast,  Figure \ref{fig:Evan:SubspaceDirectionsCorrespond} also demonstrates that $\Vh{}_3 \simeq \hat{L}_N$. To the PN order considered here, $\V_4 = \hat{V}_{h4} = \hat{L}_N$, 
since $\avL_4$ and $\avLh_4$ do not have off-diagonal terms.

\subsection{Eigenvalues}

Generally, the eigenvalues of $\avL$ are powerful diagnostics for gravitational wave signals with multiple
harmonics, particularly in the presence of precession. 
For example, to an excellent approximation, the eigenvalues of any constant-$l$ subspace orientation tensor $\avL_l$ are $l^2/2$
and $l/2$. As noted in \cite{gwastro-mergers-nr-Alignment-ROS-Methods}, the dominant eigenvalue of $\avL$, summing over many
constant-$l$ subspaces, scales as $\left<m^2\right>$ and therefore reflects the relative impact of higher-order
harmonics to the signal.

To demonstrate the value of these eigenvalues as a diagnostic, in Figure \ref{fig:Evan:Eigenvalues} we show the
\emph{relative difference} in the eigenvalues of $\avL$ and $\avLh$ from the naive leading-order predictions described
above. The PN corrections create rather small perturbations to the eigenvalues. The PN corrections within the diagonal blocks corresponding to the orbital plane and the Newtonian orbital angular momentum, such as those in Eqs.~(\ref{eq:NPfirst}) - (\ref{eq:NPlast}), can either raise or lower the eigenvalue depending on their sign. The off-diagonal PN corrections will \emph{always} increase the dominant eigenvalue, regardless of their sign. See Appendix~\ref{appB} for details.

The PN corrections of course become largest at late times, when $v$ increases rapidly, and this is why deviations in the eigenvalues grow rapidly at the end. For the $l=3$ subspace, the PN correction in the $\hat{L}_N$ block [i.e. multiplying $M^{(3)}$ in Eqs.~(\ref{eq:NP3}) and (\ref{eq:NPh3})] is always negative and hence the dominant eigenvalue is always less than its leading-order value. For the $l=2$ mode, note that the 1PN and 1.5PN corrections have the opposite sign. At early times, the 1PN term dominates and tends to decrease the dominant eigenvalue. Depending on the values of $\delta$ and ${\cal S}$, the 1.5PN correction can actually become larger than the 1PN correction and change the sign of the perturbation to the dominant eigenvalue (as happens in our BH-NS binary). 

As with the trajectory $\hat{V}$, the eigenvalue of $\avL_2$ and $\avL_3$ changes smoothly on the precession and
inspiral timescales.  By contrast, like the trajectory $\Vh$, the eigenvalues of $\avLh_2$ oscillate rapidly on the
\emph{orbital} timescale; see the bottom two panels.  [At the post-Newtonian orders available at present, the
  orientations and eigenvalues of $\avLh_3$ do not oscillate significantly.]

\section{Numerical evaluation and subtleties of self-consistent post-Newtonian expansions}

The post-Newtonian series converges slowly in powers of $v$, itself only  between  
$0.1-0.5$ for our examples. The details of how PN expressions are truncated or resummed
can quantitatively (though not qualitatively) change the conclusions described above.  
Our concern is not academic:  there exist several straightforward alternatives to this calculation which will cause $\avL$, $\avLh$ and the precession equations to effectively differ by unknown, higher PN order error terms.
In this section, we describe how these subtle issues associated with the post-Newtonian series can impact our results.

\subsection{Numerical evaluation of preferred orientation from mode coefficients}

To compute  $\avL$ and $\avLh$, we consistently expand
 both the numerator and denominator in a post-Newtonian series.  
Such a rational function form would be a completely reasonable choice to represent 
the rotation tensors. Instead, for simplicity in the expressions and their interpretation, we have chosen to re-expand the rational function as a polynomial via a Taylor series. The two agree to 1.5PN order, but differ by some unknown term at 2PN order. This is exactly analogous to the difference in TaylorT1 and TaylorT4 waveforms described in~\cite{2009PhRvD..80h4043B}, with the rational form corresponding to TaylorT1 and our polynomial expressions corresponding to TaylorT4.
 
Furthermore, a straightforward numerical implementation
 to generate  $\avLh$ will also differ slightly from the rational and polynomial expressions described above. This method would employ  \citet{gw-astro-mergers-approximations-SpinningPNHigherHarmonics} to compute
$h_{lm}$, then apply the formulae in Appendix~\ref{appA} to compute  $\avLh$ directly.
However, as $\avLh$ involves ratios of products $\propto h^2$, this direct calculation  implicitly includes many
higher-order post-Newtonian terms.
For example, in these products a $v^3$ correction to one $h_{lm}$ could multiply a $v^3$ correction to another $h_{lm'}$ to give a contribution to $\avL \propto v^6$ (3PN). This is only a partial PN correction, of course, because if the $h_{lm}$'s were complete to 3PN there would be many more terms at $v^6$.

Because of these subtleties, the values of $\avL$ and $\avLh$ computed via polynomial, rational function and direct numerical computation will differ from one another. While all of the plots in this paper were computed using the polynomial expressions of Eqs.~(\ref{eq:Pfirst})-(\ref{eq:Plast}), we have checked that the rational and direct numerical computations are quite similar. There are quantitative differences, but not qualitative ones. The differences among $\V$, $\Vh$, $\hat{L}$ and $\hat{L}_N$ are quite similar, the trajectory of $\Vh$ still has its distinctive ``crown'' shape, and our conclusions are not changed.

\subsection{Orbit-averaging the precession of $L_N$ and $L$}

Note that to create the plots in Figures \ref{fig:Evan:Trajectories:BHNS} and \ref{fig:Evan:Trajectories:BBH} we have used an orbit-averaged formula to describe the precession of $\hat{L}_N$, Eq.~(\ref{eq:prec_eq}). As a result, $\hat{L}_N$ moves on a smooth precession cone and $\hat{L}$ nutates on an orbital timescale because of its $\mathbf{L}_{SO}$ term.
This is in sharp contrast with Fig. 6 of~\cite{gwastro-mergers-nr-ComovingFrameExpansionSchmidt2010}, who find that $\hat{L}$ varies smoothly, while $\hat{L}_N$ nutates on an orbital timescale. This is presumably because the authors of~\cite{gwastro-mergers-nr-ComovingFrameExpansionSchmidt2010} used an orbit-averaged precession equation for $\hat{L}$ (rather than $\hat{L}_N$). As explained in the text immediately after Eq. (4.17) of ~\cite{1995PhRvD..52..821K}, $\hat{L}_N$ and $\hat{L}$ have the same orbit-averaged precession equation, because $\mathbf{\dot{L}}_{SO}=0$ when averaged over a circular orbit.

We emphasize that our expressions for $\avL$ and $\avLh$ in Sec.~\ref{sec:ApplyPN} do not make any sort of orbit-averaged approximation. Indeed, the use of a PN expansion for $h_{lm}$ and $\WeylScalar_{lm}$ is the only approximation involved in deriving these expressions. The orbit-averaging approximation is applied when integrating the equations of motion to generate time series for $\hat{L}_N$ and the spin vectors to plug into the expressions. Thus, the details of the plots in Sec.~\ref{sec:NumericalExamples} may vary depending on if or how orbit averaging is applied.

Therefore, when constructing PN waveforms or performing further studies to find preferred directions for precessing binaries, one should carefully consider how the precession equations handle this orbit averaging. Several authors have investigated more accurate formulations of the precession equations which do not assume orbit averaging~\cite{1995PhRvD..52..821K,2004PhRvD..70l4020S,2006PhRvD..74j4034B,2010PhRvD..81h4025G,2011PhRvD..84f4041B}. 
Given the stringent accuracy requirements needed for parameter estimation, comparison with numerical simulations and other applications, we expect that the difference between non-orbit-averaged and orbit-averaged precession equations could be significant. Therefore, we recommend future studies incorporate analytic waveforms with non-orbit-averaged precession equations to be as accurate as possible.

\ForInternalReference{
\subsection{Other possible things (*)}

- Express LL using Riemann tensor derivatives?  (Or explicitly construct 3+1-like split?) Seems wasteful/duplicative

-  Self-consistent constraint issue: because $\iota$ is the inclination of L relative to J, the inclination, v, and spin
magnitudes aren't free...they are locked in concert.  Hmm.

I don't think we need to use the coordinate frame being aligned with $J$ at all.

}

\section{Conclusions}
Gravitational wave emission from precessing compact bianries is expected to appear simpler in a suitably-chosen frame.  To leading post-Newtonian order, this frame
should correspond to the orbital angular momentum at large separations.  Late in the inspiral, many plausible reference
orientations exist.  In this paper, we have demonstrated that one choice for that orientation (the prinicipal
eigendirection of $\avL$) does not agree with either of two others (the orbital and total angular momenta).

The suitability of this or any other reference frame is best judged by whether it facilitates useful insights and
calculations.  
In this paper we demonstrate the tensor $\avL$ provides a particularly tractable way to extract preferred orientations
from asymptotic radiation.   Our approach   allows different preferred orientations to exist at each
multipolar order.   Previous papers have demonstrated this method useful in analyzing numerical relativity
simulations \cite{gwastro-mergers-nr-Alignment-ROS-Methods,gwastro-mergers-nr-Alignment-ROS-IsJEnough}
Finally, in the text and appendix we provide methods to evaluate this average  using standard tables of $h_{lm}$
modes.
\hidetosubmit{s as well as the underlying STF radiative multipoles.}

Our investigations suggest the proposed orientations will differ from one another for \emph{any} precessing binary, regardless of its mass ratio or spin orientation. Furthermore, we note that in the PN regime the behavior of these preferred directions can be noticeably affected by the details of how the PN series is truncated or resummed and also by the manner in which orbit averaging of the precession equations is performed. This suggests that precessing PN models could benefit greatly from using more general precession equations which do not assume orbit averaging.

\begin{acknowledgements}
EO and ROS are supported by NSF award PHY-0970074.  ROS is also supported by the Bradley Program Fellowship and the UWM Research
Growth Initiative. 
\end{acknowledgements}

\appendix
\section{Evaluating the average} \label{appA}
For reference, we provide explicit formula that relate the mode amplitudes $\psi_{lm} $ of the Weyl scalar to the
orientation-averaged expression $\avL$.   We first calculate two real and two complex quantities: 
\begin{eqnarray}
I_2&\equiv&  \frac{1}{2}\,(\psi,L_+L_+\psi) \nonumber \\
 &=& \frac{1}{2}\,\sum_{lm} c_{lm}c_{l,m+1} \psi_{l,m+2}^*\psi_{lm} \\
I_1 & \equiv &(\psi,L_+(L_z+1/2)\psi)  \nonumber \\
 &=& \sum_{lm} c_{lm}(m+1/2) \psi_{l,m+1}^*\psi_{lm} \\
I_0 &\equiv& \frac{1}{2}\left(\psi| L^2 - L_z^2 |\psi\right) \nonumber\\
 &=& \frac{1}{2}\sum_{lm} [l(l+1)-m^2]|\psi_{lm}|^2  \\
I_{zz} &\equiv& (\psi,L_z L_z \psi) = \sum_{lm} m^2 |\psi_{lm}|^2 
\end{eqnarray}
where $c_{lm} = \sqrt{l(l+1)-m(m+1)}$.
In terms of these expressions, the orientation-averaged tensor is
\begin{eqnarray}
\avL &=& 
\begin{bmatrix}
 I_0 + \text{Re}(I_2) & \text{Im} I_2  &   \text{Re} I_1 \\
  &   I_0 - \text{Re}(I_2) & \text{Im} I_1 \\
 & & I_{zz}
\end{bmatrix}
\end{eqnarray}

In a frame aligned with $L$, the orbital angular momentum ($\propto \hat{z}$) will be the principal axis of $\avL$  if
and only if  $I_1  = 0$.   %
The  prefactor $C_{lm}\equiv c_{lm}(m+1/2)$ in the definition of $I_1$ is antisymmetric about $m=-1/2$ -- that is, under a
transformation $m\rightarrow -(m+1)$.  
Therefore,  one way to make  $I_1=0$  and thus $L$ a principal axis of $\avL$ occurs if $\psi_{lm+1}^*\psi_{lm}$ is
\emph{symmetric} about $m=-1/2$.  
For nonprecessing binaries the strain modes satisfy $h_{lm}^*=h_{l,-m}$ (reflection symmetry through the $xy$ plane) and evolve in
phase with the orbit ($h_{lm}= e^{i m \Psi} A_{l,|m|}$ for some real $A_{lm}$).  Therefore, for each constant-$l$ subspace,
term-by-term cancellation occurs in $I_1$, independent of the details of $A_{lm}$
As a concrete example, we expand out $I_1$ for $l=2$ as 
\begin{eqnarray}
I_1 &=& e^{i \Psi} [ A_{20} A_{21} (C_{2-1}+C_{20}) + A_{21}A_{22}(C_{2-2}+C_{2,1})] \nonumber \\ &=&0
\end{eqnarray}

\section{Perturbations of the eigensystems} \label{appB}

First, let us consider a non-precessing binary and study $\avL_l$ in a frame aligned with $L_N$. From Eq.~(\ref{eq:tauL}), we see that to leading-order
\begin{eqnarray}
\avL_l \simeq {\cal T}^{(l)} =  \left( \begin{matrix}
l/2 & 0 & 0\\
0 & l/2 & 0\\
0 & 0 & l^2
\end{matrix}\right)\ ,
\end{eqnarray}
and of course the eigenvalues are $l^2$, $l/2$ and $l/2$. The PN corrections are small perturbations to the tensor, so when we add them in it takes the form
\begin{widetext}
\begin{eqnarray}
\avL_l = {\cal T}^{(l)} + \epsilon_1 M^{(1)} + \epsilon_2 M^{(2)} +  \epsilon_3 M^{(3)}  =  \left( \begin{matrix}
l/2 + \epsilon_1 + \epsilon_2 \cos 2 \Phi & \epsilon_2 \sin 2 \Phi & 0\\
\epsilon_2 \sin 2 \Phi & l/2 + \epsilon_1 - \epsilon_2 \cos 2\Phi & 0\\
0 & 0 & l^2 + \epsilon_3
\end{matrix}\right)\ .
\end{eqnarray}
\end{widetext}
In this case, the eigenvalues become $l^2 + \epsilon_3$ and $l/2 + \epsilon_1 \pm \epsilon_2$. So, PN corrections to the diagonal blocks (corresponding to either the orbital plane or the Newtonian orbital angular momentum) will linearly perturb the eigenvalues within that block. 
They do not change the dominant eigenvector.   
While they do weakly break degeneracy in the orbital plane, setting two distinct eigenvectors rotating in the orbital plane, 
they do so weakly; the space spanned by these two vectors is still the subspace of the unperturbed system.

Now let us consider the perturbations from the off-diagonal terms that are present in precessing binaries. For simplicity, we will consider the $l=2$ case, although the same argument can be applied to any value of $l$ and it is straightforward to derive more general expressions. In this case, in a frame aligned with the instantaneous $\hat{L}_N$, our perturbed rotation tensor will have the form
\begin{equation}
\avL_2 =  \left( \begin{matrix}
1 & 0 & \epsilon_1\\
0 & 1 & \epsilon_2\\
\epsilon_1 & \epsilon_2 & 4
\end{matrix}\right)\ .
\end{equation}
The eigenvalues of this system are 
\begin{eqnarray}
\frac{1}{2}(5 + 3 \sqrt{1 + \frac{4}{9}(\epsilon_1^2 + \epsilon_2^2)}) &\simeq& 4 + \frac{\epsilon_1^2 + \epsilon_2^2}{3}\\
\frac{1}{2} (5 - 3 \sqrt{1 + \frac{4}{9}(\epsilon_1^2 + \epsilon_2^2)}) &\simeq& 1 +  \frac{\epsilon_1^2 + \epsilon_2^2}{3}\\
  1 &&
 \end{eqnarray}
In particular, we see that the presence of these off-diagonal terms will \emph{always} increase the magnitude of the dominant eigenvalue, regardless of the sign of these perturbations. It is also worth noting that the dominant (unnormalized) eigenvector is given by
\begin{equation}
V_2 = \left\{ \epsilon_1, \epsilon_2, \frac{3}{2} \left( 1 + \sqrt{1 + \frac{4}{9}(\epsilon_1^2 + \epsilon_2^2)} \right) \right\}\ ,
\end{equation}
so that $V_{2x}$ ($V_{2y}$) is linearly proportional to $\left< {\cal L}_{(x} {\cal L}_{z)} \right>_2$ ($\left< {\cal L}_{(y} {\cal L}_{z)} \right>_2$). This shows quite clearly, for example, the fact that oscillatory off-diagonal terms such as those of the form in Eq.~(\ref{eq:STF5}) are directly responsible for oscillations in the principal eigendirections.

\bibliography{paperexport}

\begin{thebibliography}{22}
\expandafter\ifx\csname natexlab\endcsname\relax\def\natexlab#1{#1}\fi
\expandafter\ifx\csname bibnamefont\endcsname\relax
  \def\bibnamefont#1{#1}\fi
\expandafter\ifx\csname bibfnamefont\endcsname\relax
  \def\bibfnamefont#1{#1}\fi
\expandafter\ifx\csname citenamefont\endcsname\relax
  \def\citenamefont#1{#1}\fi
\expandafter\ifx\csname url\endcsname\relax
  \def\url#1{\texttt{#1}}\fi
\expandafter\ifx\csname urlprefix\endcsname\relax\def\urlprefix{URL }\fi
\providecommand{\bibinfo}[2]{#2}
\providecommand{\eprint}[2][]{\url{#2}}

\bibitem[{\citenamefont{Apostolatos et~al.}(1994)\citenamefont{Apostolatos,
  Cutler, Sussman, and Thorne}}]{ACST}
\bibinfo{author}{\bibfnamefont{T.~A.} \bibnamefont{Apostolatos}},
  \bibinfo{author}{\bibfnamefont{C.}~\bibnamefont{Cutler}},
  \bibinfo{author}{\bibfnamefont{G.~J.} \bibnamefont{Sussman}},
  \bibnamefont{and} \bibinfo{author}{\bibfnamefont{K.~S.}
  \bibnamefont{Thorne}}, \bibinfo{journal}{\prd} \textbf{\bibinfo{volume}{49}},
  \bibinfo{pages}{6274} (\bibinfo{year}{1994}).

\bibitem[{\citenamefont{{Schmidt} et~al.}(2011)\citenamefont{{Schmidt},
  {Hannam}, {Husa}, and
  {Ajith}}}]{gwastro-mergers-nr-ComovingFrameExpansionSchmidt2010}
\bibinfo{author}{\bibfnamefont{P.}~\bibnamefont{{Schmidt}}},
  \bibinfo{author}{\bibfnamefont{M.}~\bibnamefont{{Hannam}}},
  \bibinfo{author}{\bibfnamefont{S.}~\bibnamefont{{Husa}}}, \bibnamefont{and}
  \bibinfo{author}{\bibfnamefont{P.}~\bibnamefont{{Ajith}}},
  \bibinfo{journal}{\prd} \textbf{\bibinfo{volume}{84}}, \bibinfo{eid}{024046}
  (\bibinfo{year}{2011}),
  \urlprefix\url{http://xxx.lanl.gov/abs/arXiv:1012.2879}.

\bibitem[{\citenamefont{{Arun} et~al.}(2009)\citenamefont{{Arun}, {Buonanno},
  {Faye}, and
  {Ochsner}}}]{gw-astro-mergers-approximations-SpinningPNHigherHarmonics}
\bibinfo{author}{\bibfnamefont{K.~G.} \bibnamefont{{Arun}}},
  \bibinfo{author}{\bibfnamefont{A.}~\bibnamefont{{Buonanno}}},
  \bibinfo{author}{\bibfnamefont{G.}~\bibnamefont{{Faye}}}, \bibnamefont{and}
  \bibinfo{author}{\bibfnamefont{E.}~\bibnamefont{{Ochsner}}},
  \bibinfo{journal}{\prd} \textbf{\bibinfo{volume}{79}},
  \bibinfo{pages}{104023} (\bibinfo{year}{2009}).

\bibitem[{\citenamefont{{ Brown} et~al.}(2012)\citenamefont{{ Brown},
  {Lundgren}, and
  {O'Shaughnessy}}}]{gw-astro-SpinAlignedLundgren-FragmentA-Theory}
\bibinfo{author}{\bibfnamefont{D.}~\bibnamefont{{ Brown}}},
  \bibinfo{author}{\bibfnamefont{A.}~\bibnamefont{{Lundgren}}},
  \bibnamefont{and}
  \bibinfo{author}{\bibfnamefont{R.}~\bibnamefont{{O'Shaughnessy}}},
  \bibinfo{journal}{Submitted to PRD (arXiv:1203.6060)}
  (\bibinfo{year}{2012}), \urlprefix\url{http://arxiv.org/abs/1203.6060}.

\bibitem[{\citenamefont{{Boyle} et~al.}(2011)\citenamefont{{Boyle}, {Owen}, and
  {Pfeiffer}}}]{gwastro-mergers-nr-Alignment-BoyleHarald-2011}
\bibinfo{author}{\bibfnamefont{M.}~\bibnamefont{{Boyle}}},
  \bibinfo{author}{\bibfnamefont{R.}~\bibnamefont{{Owen}}}, \bibnamefont{and}
  \bibinfo{author}{\bibfnamefont{H.~P.} \bibnamefont{{Pfeiffer}}},
  \bibinfo{journal}{\prd} \textbf{\bibinfo{volume}{84}}, \bibinfo{eid}{124011}
  (\bibinfo{year}{2011}).

\bibitem[{\citenamefont{{O'Shaughnessy}
  et~al.}(2011)\citenamefont{{O'Shaughnessy}, {Vaishnav}, {Healy}, {Meeks}, and
  {Shoemaker}}}]{gwastro-mergers-nr-Alignment-ROS-Methods}
\bibinfo{author}{\bibfnamefont{R.}~\bibnamefont{{O'Shaughnessy}}},
  \bibinfo{author}{\bibfnamefont{B.}~\bibnamefont{{Vaishnav}}},
  \bibinfo{author}{\bibfnamefont{J.}~\bibnamefont{{Healy}}},
  \bibinfo{author}{\bibfnamefont{Z.}~\bibnamefont{{Meeks}}}, \bibnamefont{and}
  \bibinfo{author}{\bibfnamefont{D.}~\bibnamefont{{Shoemaker}}},
  \bibinfo{journal}{\prd} \textbf{\bibinfo{volume}{84}},
  \bibinfo{pages}{124002} (\bibinfo{year}{2011}),
  \urlprefix\url{http://link.aps.org/doi/10.1103/PhysRevD.84.124002}.

\bibitem[{\citenamefont{{Kidder}}(1995)}]{1995PhRvD..52..821K}
\bibinfo{author}{\bibfnamefont{L.~E.} \bibnamefont{{Kidder}}},
  \bibinfo{journal}{\prd} \textbf{\bibinfo{volume}{52}}, \bibinfo{pages}{821}
  (\bibinfo{year}{1995}).

\bibitem[{\citenamefont{{Blanchet} et~al.}(2008)\citenamefont{{Blanchet},
  {Faye}, {Iyer}, and {Sinha}}}]{2008CQGra..25p5003B}
\bibinfo{author}{\bibfnamefont{L.}~\bibnamefont{{Blanchet}}},
  \bibinfo{author}{\bibfnamefont{G.}~\bibnamefont{{Faye}}},
  \bibinfo{author}{\bibfnamefont{B.~R.} \bibnamefont{{Iyer}}},
  \bibnamefont{and} \bibinfo{author}{\bibfnamefont{S.}~\bibnamefont{{Sinha}}},
  \bibinfo{journal}{Classical and Quantum Gravity}
  \textbf{\bibinfo{volume}{25}}, \bibinfo{pages}{165003}
  (\bibinfo{year}{2008}).

\bibitem[{\citenamefont{{Kidder}}(2008)}]{gwastro-pn-MultipoleMomentsNonspinning}
\bibinfo{author}{\bibfnamefont{L.~E.} \bibnamefont{{Kidder}}},
  \bibinfo{journal}{\prd} \textbf{\bibinfo{volume}{77}},
  \bibinfo{pages}{044016} (\bibinfo{year}{2008}).

\bibitem[{\citenamefont{Will and Wiseman}(1996)}]{WillWiseman:1996}
\bibinfo{author}{\bibfnamefont{C.~M.} \bibnamefont{Will}} \bibnamefont{and}
  \bibinfo{author}{\bibfnamefont{A.~G.} \bibnamefont{Wiseman}},
  \textbf{\bibinfo{volume}{54}}, \bibinfo{pages}{4813} (\bibinfo{year}{1996}).

\bibitem[{\citenamefont{{Faye} et~al.}(2006)\citenamefont{{Faye}, {Blanchet},
  and {Buonanno}}}]{2006PhRvD..74j4033F}
\bibinfo{author}{\bibfnamefont{G.}~\bibnamefont{{Faye}}},
  \bibinfo{author}{\bibfnamefont{L.}~\bibnamefont{{Blanchet}}},
  \bibnamefont{and}
  \bibinfo{author}{\bibfnamefont{A.}~\bibnamefont{{Buonanno}}},
  \bibinfo{journal}{\prd} \textbf{\bibinfo{volume}{74}},
  \bibinfo{pages}{104033} (\bibinfo{year}{2006}).

\bibitem[{\citenamefont{{Will}}(2005)}]{2005PhRvD..71h4027W}
\bibinfo{author}{\bibfnamefont{C.~M.} \bibnamefont{{Will}}},
  \bibinfo{journal}{\prd} \textbf{\bibinfo{volume}{71}}, \bibinfo{eid}{084027}
  (\bibinfo{year}{2005}).

\bibitem[{\citenamefont{{Wiseman} and {Will}}(1991)}]{1991PhRvD..44.2945W}
\bibinfo{author}{\bibfnamefont{A.~G.} \bibnamefont{{Wiseman}}}
  \bibnamefont{and} \bibinfo{author}{\bibfnamefont{C.~M.}
  \bibnamefont{{Will}}}, \bibinfo{journal}{\prd} \textbf{\bibinfo{volume}{44}},
  \bibinfo{pages}{2945} (\bibinfo{year}{1991}).

\bibitem[{\citenamefont{{Favata}}(2010)}]{2010CQGra..27h4036F}
\bibinfo{author}{\bibfnamefont{M.}~\bibnamefont{{Favata}}},
  \bibinfo{journal}{Classical and Quantum Gravity}
  \textbf{\bibinfo{volume}{27}}, \bibinfo{pages}{084036}
  (\bibinfo{year}{2010}).

\bibitem[{\citenamefont{{LIGO Scientific Collaboration}}(2009)}]{LAL}
\bibinfo{author}{\bibnamefont{{LIGO Scientific Collaboration}}},
  \emph{\bibinfo{title}{{\normalfont LSC Algorithm Library}}}
  (\bibinfo{year}{2009}),
  \urlprefix\url{http://www.lsc-group.phys.uwm.edu/lal}.

\bibitem[{\citenamefont{{Pan} et~al.}(2004)\citenamefont{{Pan}, {Buonanno},
  {Chen}, and {Vallisneri}}}]{BCV:PTF}
\bibinfo{author}{\bibfnamefont{Y.}~\bibnamefont{{Pan}}},
  \bibinfo{author}{\bibfnamefont{A.}~\bibnamefont{{Buonanno}}},
  \bibinfo{author}{\bibfnamefont{Y.}~\bibnamefont{{Chen}}}, \bibnamefont{and}
  \bibinfo{author}{\bibfnamefont{M.}~\bibnamefont{{Vallisneri}}},
  \bibinfo{journal}{\prd} \textbf{\bibinfo{volume}{69}},
  \bibinfo{pages}{104017} (\bibinfo{year}{2004}),
  \urlprefix\url{http://xxx.lanl.gov/abs/gr-qc/0310034}.

\bibitem[{\citenamefont{{Schnittman}}(2004)}]{2004PhRvD..70l4020S}
\bibinfo{author}{\bibfnamefont{J.~D.} \bibnamefont{{Schnittman}}},
  \bibinfo{journal}{\prd} \textbf{\bibinfo{volume}{70}},
  \bibinfo{pages}{124020} (\bibinfo{year}{2004}).

\bibitem[{\citenamefont{{Buonanno} et~al.}(2009)\citenamefont{{Buonanno},
  {Iyer}, {Ochsner}, {Pan}, and {Sathyaprakash}}}]{2009PhRvD..80h4043B}
\bibinfo{author}{\bibfnamefont{A.}~\bibnamefont{{Buonanno}}},
  \bibinfo{author}{\bibfnamefont{B.~R.} \bibnamefont{{Iyer}}},
  \bibinfo{author}{\bibfnamefont{E.}~\bibnamefont{{Ochsner}}},
  \bibinfo{author}{\bibfnamefont{Y.}~\bibnamefont{{Pan}}}, \bibnamefont{and}
  \bibinfo{author}{\bibfnamefont{B.~S.} \bibnamefont{{Sathyaprakash}}},
  \bibinfo{journal}{\prd} \textbf{\bibinfo{volume}{80}}, \bibinfo{eid}{084043}
  (\bibinfo{year}{2009}), \eprint{0907.0700}.

\bibitem[{\citenamefont{{O'Shaughnessy}
  et~al.}(2012)\citenamefont{{O'Shaughnessy}, {Healy}, {London}, {Meeks}, and
  {Shoemaker}}}]{gwastro-mergers-nr-Alignment-ROS-IsJEnough}
\bibinfo{author}{\bibfnamefont{R.}~\bibnamefont{{O'Shaughnessy}}},
  \bibinfo{author}{\bibfnamefont{J.}~\bibnamefont{{Healy}}},
  \bibinfo{author}{\bibfnamefont{L.}~\bibnamefont{{London}}},
  \bibinfo{author}{\bibfnamefont{Z.}~\bibnamefont{{Meeks}}}, \bibnamefont{and}
  \bibinfo{author}{\bibfnamefont{D.}~\bibnamefont{{Shoemaker}}},
  \bibinfo{journal}{PRD in press (arXiv:1201.2113)}  (\bibinfo{year}{2012}),
  \urlprefix\url{http://xxx.lanl.gov/abs/1201.2113}.

\bibitem[{\citenamefont{{Blanchet} et~al.}(2006)\citenamefont{{Blanchet},
  {Buonanno}, and {Faye}}}]{2006PhRvD..74j4034B}
\bibinfo{author}{\bibfnamefont{L.}~\bibnamefont{{Blanchet}}},
  \bibinfo{author}{\bibfnamefont{A.}~\bibnamefont{{Buonanno}}},
  \bibnamefont{and} \bibinfo{author}{\bibfnamefont{G.}~\bibnamefont{{Faye}}},
  \bibinfo{journal}{\prd} \textbf{\bibinfo{volume}{74}},
  \bibinfo{pages}{104034} (\bibinfo{year}{2006}).

\bibitem[{\citenamefont{{Gergely}}(2010)}]{2010PhRvD..81h4025G}
\bibinfo{author}{\bibfnamefont{L.~{\'A}.} \bibnamefont{{Gergely}}},
  \bibinfo{journal}{\prd} \textbf{\bibinfo{volume}{81}},
  \bibinfo{pages}{084025} (\bibinfo{year}{2010}).

\bibitem[{\citenamefont{{Blanchet} et~al.}(2011)\citenamefont{{Blanchet},
  {Buonanno}, and {Faye}}}]{2011PhRvD..84f4041B}
\bibinfo{author}{\bibfnamefont{L.}~\bibnamefont{{Blanchet}}},
  \bibinfo{author}{\bibfnamefont{A.}~\bibnamefont{{Buonanno}}},
  \bibnamefont{and} \bibinfo{author}{\bibfnamefont{G.}~\bibnamefont{{Faye}}},
  \bibinfo{journal}{\prd} \textbf{\bibinfo{volume}{84}}, \bibinfo{eid}{064041}
  (\bibinfo{year}{2011}), \eprint{1104.5659}.

\end{thebibliography}
\end{document}